\documentclass[manuscript]{emulateapj}
\usepackage{natbib}
\usepackage{epsfig}
\usepackage{lscape}
\usepackage{rotating}
\usepackage{lscape,graphicx}

\usepackage{color}

\newcommand{\cmjj}{\mbox{${\rm cm^{-2}}$}}
\newcommand{\hI}{\mbox{${\rm H\,I}$}}

\newcommand{\apg}{\gtrsim}
\newcommand{\apll}{\lesssim}
\newcommand{\etal}{\ensuremath{\mbox{et~al.}}}
\newcommand{\hmsol}{\mbox{$h^{-1}\,{\rm M}_\odot$}}

\slugcomment{Submitted to the Astrophysical Journal}

\shorttitle{MgII \& LRGs at $z \sim 0.5$}
\shortauthors{Gauthier et al.}

\begin{document}

\title{The Clustering of Mg\,II Absorption Systems at $z \sim 0.5$ and \\ detection of cold gas in massive halos}

\author{Jean-Ren\'e Gauthier \altaffilmark{1},  Hsiao-Wen Chen \altaffilmark{1} and Jeremy L. Tinker \altaffilmark{1,2}}

\altaffiltext{1}{Department of Astronomy \& Astrophysics and Kavli Institute for Cosmological Physics, University of Chicago, IL}
\altaffiltext{2}{Berkeley Center for Cosmological Physics, University of California, Berkeley, CA}

\begin{abstract}

  We measure the large-scale clustering of Mg\,II $\lambda\lambda$
  2796,2803 absorbers with respect to a population of luminous red
  galaxies (LRGs) at $z \sim 0.5$.  From the cross-correlation
  measurements between Mg\,II absorbers and LRGs, we calculate the
  mean bias of the dark matter halos in which the absorbers reside.
  We investigate possible systematic uncertainties in the clustering
  measurements due to the sample selection of LRGs and due to
  uncertainties in photometric redshifts.  First, we compare the
  cross-correlation amplitudes determined using a {\it flux-limited}
  LRG sample and a {\it volume-limited} one.  The comparison shows
  that the relative halo bias of Mg\,II absorbers using a {\it
    flux-limited} LRG sample can be {\it overestimated} by as much as
  $\approx$ 20\%.  Next, we assess the systematic uncertainty due to
  photometric redshift errors using a mock galaxy catalog with added
  redshift uncertainties comparable to the data.  We show that the
  relative clustering amplitude measured without accounting for
  photometric redshift uncertainties is {\it overestimated} by
  $\approx 10$\%.  After accounting for these two main uncertainties,
  we find a 1-$\sigma$ anti-correlation between mean halo bias and
  absorber strength $W_r(2796)$ that translates into a 1-$\sigma$
  anti-correlation between mean galaxy mass and $W_r(2796)$. The results indicate 
  that a significant fraction of the Mg\,II
  absorber population of $W_r(2796)=1-1.5$ \AA\ are found in group-size dark
  matter halos of $\log\,M_h < 13.4$, whereas absorbers of
  $W_r(2796)>1.5$ \AA\ are primarily seen in halos of $\log\,M_h
  <12.7$. A larger dataset would improve the precision of both the 
  clustering measurements and the relationship between equivalent width and halo mass. 
  Finally, the strong clustering of Mg\,II absorbers down to
  scales of $\sim 0.3\ h^{-1}$ Mpc indicates the presence of cool gas
  inside the virial radii of the dark matter halos hosting the LRGs.
\end{abstract}


\section{Introduction}

Characterizing the structure and evolution of the cold gas in dark
matter halos is a key element in current models of galaxy formation
(e.g.,\ Kere{\v s} \etal\ 2005). The fraction of cold and hot gas within
dark matter halos and the rate at which gas is being accreted are
essential to our understanding of disc and star formation
(e.g.,\ \citealt{dekel2009a}).  Extended gaseous envelopes around
galaxies were first predicted several decades ago
(\citealt{spitzer1956a}).  Observations of H\,I maps around local
galaxies (e.g.,\ \citealt{thilker2004a,doyle2005a}) and comparisons of
galaxies and QSO absorption-line systems
(e.g.,\ \citealt{bergeron1986a,lanzetta1990a,steidel1994a,chen2001b,chen2008a})
have indeed shown the presence of extended cool gas ($T\sim 10^4$ K)
out to $50-100\ h^{-1}$ kpc radii.  The physical mechanism that
explains the origin of the extended cold halo gas is, however,
unclear.  Some of the most common scenarios are (i) outflows from starburst 
systems (e.g.,\ \citealt{bond2001a});  (ii) stripping from the
accretion of gas-rich satellites \citep{wang1993a}; (iii) cold gas
bound to substructure within the host dark halo (e.g.,\ \citealt{sternberg2002a}), 
and (iv) a two-phase medium composed of cold and hot
gas \citep{mo1996a,maller2004a}.

A potential probe of the cold halo gas is the Mg\,II $\lambda\lambda$
2796,2803 absorption features commonly seen in the spectra of
background QSOs.  These absorbers are thought to originate in
photo-ionized gas of temperature $T\sim 10^4$ K and to trace
high-column density \hI\ clouds of neutral hydrogen column density
$N(\hI) \approx 10^{18}-10^{22}$ \cmjj\
\citep{bergeron1986a,rao2006a}.  This large associated \hI\ column
density indicates that Mg\,II absorbers arise in halo gas around
individual galaxies \citep{doyle2005a}.  This is also supported by the
presence of luminous galaxies at projected distances $\rho = 50-100 \
h^{-1}$ kpc from known Mg\,II absorbers \citep{bergeron1986b,
  lanzetta1990a,bergeron1991a,lanzetta1992a,steidel1992b,steidel1994a,zibetti2007a,nestor2007a,kacprzak2008a}.

In addition to the classical gas accretion scenario for the origin of
Mg\,II absorbers at larger galactic radii, there is a competing
scenario that has gained substantial attention recently. In this new
picture, strong Mg\,II absorbers of rest-frame absorption equivalent
width $W_r(2796)>1$ \AA\ originate in starburst driven outflows
(e.g.,\ \citealt{bond2001a,menard2008a,weiner2008a}).  Under this scenario, the
Mg\,II absorbing gas orginates in the cold \emph{outflowing} material
surrounding starburst galaxies. An interesting recent finding is a
strong correlation between dust extinction $E(B-V)$ and $W_r(2796)$
by \citet{menard2008b}.  While the observed $E(B-V)$ vs.\ $W_r(2796)$
is consistent with the expectation of the starburst scenario, this
observation is also expected if the Mg\,II absorbing galaxies exhibit
a metallicity gradient commonly seen in regular galaxies
(e.g.,\ \citealt{zaritsky1994a,vanzee1998a}).  Dense clumps in starburst
driven outflows are expected to contribute to some fraction of the
observed Mg\,II absorbers, but the significance of this fraction and
how the fraction varies with $W_r(2796)$ are both uncertain.

As a first step toward a quantitative understanding of the physical
origin of the Mg\,II absorber population, we are carrying out a
cross-correlation analysis of Mg\,II absorbers with photometrically
identified luminous red galaxies (LRGs) in the Sloan Digital Sky
Survey (SDSS; \citealt{york2000a}).  The primary goals are (1) to
determine the clustering amplitude of Mg\,II absorbers and (2) to
examine how the clustering amplitude depends on absorber strength
$W_r(2796)$.  The clustering amplitude of Mg\,II absorbers is
determined based on their cross-correlation signals with LRGs on
projected co-moving distance scales of $r_p=1-30\ h^{-1}$ Mpc.  Because the mean
halo\footnote{We define halo as a region of overdensity 200 with
  respect to the mean mass density in the universe.} mass of the LRGs
can be calculated from their clustering signal
(e.g.,\ \citealt{zheng2008a,blake2008a,wake2008a,padmanabhan2008a}) the cross-correlation amplitude of LRGs
and Mg\,II absorbers provides a statistical estimate of the mean mass
of the host dark matter halos.  A similar study was published by
Bouch\'e \etal\ (2004, 2006), in which the authors attempted to
constrain the mean halo mass of Mg\,II absorbers using a flux-limited
($i'<21$ mag) sample of LRGs found in SDSS (see also Lundgren \etal\
2009).

Our analysis differs from others in two important aspects.  First, we
measure the clustering amplitude of Mg\,II absorbers using a {\it
  volume-limited} (instead of flux-limited) LRG sample.  A
flux-limited selection criterion forms an inhomogeneous sample of
LRGs, excluding progressively more intrinsically fainter LRGs (and
hence lower-mass halos) toward higher redshifts.  Such inhomogeneous
samples of LRGs over a broad redshift range ($z=0.35-0.8$) include an
inherent uncertainty in the estimated mean halo mass that is difficult
to assess, but this systematic bias has not been addressed in previous
studies.  We have directly compared the clustering amplitudes between
using a flux-limited and a volume-limited LRG sample.  We will show in
the following sections that the relative halo bias of Mg\,II absorbers
may have been {\it overestimated} by as much as $\approx$ 20\% in
previous studies.

Second, the LRGs in SDSS have been identified using photometric
redshifts ($z_{\rm phot}$) that have associated uncertainties relative
to spectroscopic redshifts ($z_{\rm spec}$) of $\sigma_0=|z_{\rm
  phot}-z_{\rm spec}|/(1+z_{\rm spec})\approx0.03$ at $i'\approx 19$
(\citealt{collister2007a}; \citealt{oyaizu2008a}).  At fainter
magnitudes, $\sigma_0$ increases steeply to $\sigma_0\ge 0.1$ at
$i'>20.7$ (Collister \etal\ 2007).  It is clear from these studies
that the redshift uncertainties of the LRG sample vary from
$\sigma_z=\sigma_0(1+z)=0.04$ for galaxies of $i'<19$ at
$z=0.35$ to $\sigma_z=0.18$ for galaxies of $i'\approx 21$ at $z=0.8$.
On the other hand, Mg\,II absorbers are identified in QSO spectra with
a redshift precision better than $\sigma_z \approx 0.0004$, corresponding
to roughly half of the width of a resolution element. The
dependence of photometric redshift errors on galaxy brightness and
redshift are expected to introduce additional systematic uncertainties
in the estimate of the Mg\,II-LRG cross-correlation amplitude.  To
reduce the systematic bias due to photometric redshift errors, we
first restrict our analysis to including only galaxies brighter than
$i'=20$\footnote{The volume-limited selection criterion is applied
  after adopting this brightness limit.}.  This allows us to maintain
an LRG sample of higher redshift precisions.  In addition, we assess
the systematic uncertainty using a mock galaxy catalog with and
without an added $\sigma_0\times (1+z)$ redshift perturbation.  We
will show in the following section that relative clustering amplitude
of Mg\,II absorbers measured without accounting for photometric
redshift uncertainties may have been {\it overestimated} by $\approx
10$\%.

This paper is organized as follows. We first describe our data
samples, Mg\,II and LRGs in Section 2. In Section 3.1, we describe the
method used to calculate the two-point correlation function and
associated errors. 
Cross-correalation results are presented in section 3.2, including a
discussion of the effects of photometric redshift uncertainties in the
calculation.  In Section 4, we summarize the routines adopted for 
calculating the bias and mean halo mass of absorbers.  
Finally, we discuss the results of our analysis in Section 5.
A more extensive halo occupation distribution analysis of the Mg\,II
absorbers will be the subject of a forthcoming paper.  We adopt a
$\Lambda$CDM cosmology, $\Omega_{\rm M}=0.25$ and $\Omega_\Lambda =
0.75$, with a dimensionless Hubble parameter $h = H_0/(100 \ {\rm km} \
{\rm s}^{-1}\ {\rm Mpc}^{-1})$ throughout the paper. All masses are expressed in units 
of $h^{-1}~M_{\odot}$ unless otherwise stated. 

\section{Data}

\subsection{MgII Catalog}
The MgII absorbers catalog is based on an extension of the SDSS DR3
sample \citep{prochter2006a} to include DR5 QSO spectra (J.X.\
Prochaska, private communication). This sample of Mg\,II absorbers has
a 95\% completeness for absorbers of $W_r(2796) > 1$
\AA. The catalog
contains 11,254 Mg\,II absorbers detected at $z_{\rm Mg\,II}=0.37-2.3$ along 9,774 QSO
sightlines. From this sample, we selected absorbers with a separation
of at least 10,000 km/s from the QSO redshift to avoid contamination
by associated absorbers of the QSOs.  We excluded absorbers that fall
outside of our survey mask (see Section 3.1.1) and we limited ourselves
to Mg\,II doublets in the redshift range of interest, $z=0.40-0.70$,
defined by the LRG sample.  Note that none of the absorbers listed in
the Prochter et al.\ catalog are found in the Ly$\alpha$ forest of
their respective quasar.  The procedure described above yielded 1,158
absorbers of $W_r(2796) > 1$ \AA.  This contitutes our \emph{primary}
Mg\,II catalog. The short dashed histogram in Figure \ref{lrgs_z_catalog}
shows the redshift distribution of our primary Mg\,II sample.



\subsection{LRG Catalog}
LRGs are tracers of the large-scale structures in the universe. 
Their clustering properties are well-known, and they reside in massive halos of
$M_h>10^{13} \hmsol$ (e.g.,\ \citealt{blake2008a,zheng2008a,wake2008a,padmanabhan2008a}). 
Because they are luminous and have strong 4,000-\AA\ break, their photometric redshift can be more
reliably estimated than blue star-forming galaxies.

We used the MegaZ-LRG catalog (hereafter MegaZ) \citep{collister2007a}
as our initial LRG sample.  MegaZ is a photometric redshift catalog of
approximately $10^6$ LRGs found in the SDSS DR4 imaging footprint.
The catalog covers more than 5,000 $\rm{deg}^2$ in the redshift range
0.4 $<$ $z$ $<$ 0.7.  The LRGs are selected following a series of cuts
in a multidimensional color diagram (e.g.,\ \citealt{scranton2003a,eisenstein2005a,collister2007a}).
The photometric redshifts of
LRGs are determined using artificial neural networks (ANNz) and the
2SLAQ training set of 13,000 available LRGs spectroscopic redshifts
\citep{cannon2006a}.

We applied a set of additional color selection criteria suggested by
\citet{blake2007a}.  These modified criteria yielded less
contamination by blue galaxies and were used in clustering and halo
occupation analyses \citep{blake2008a}.  We further limited ourselves 
to galaxies with $i'<20$ for a higher photometric redshift precision. 
We defined three LRG samples for our analysis.  The first was a flux-limited sample of $i'<20$ LRGs at $z=0.4-0.7$.  This primary sample covers the entire redshift range offered by the initial LRG sample.  The second was a volume-limited sample of $M_{i'}-5\,\log\,h<-22$ at
$z=0.45-0.60$.  The redshift range was selected to provide the largest number of LRGs available under a uniform minimum rest-frame absolute magnitude selection criterion.  Extending to lower or higher redshifts would result in a significant reduction in the sample size.  The third was 
a flux-limited sample of $i'<20$ at $z=0.45-0.60$ to be directly compared with the volume-limited subsample from the same redshift range.  

Our sample definition was motivated by the knowledge that a flux-limited sample contains galaxies that are progressively fainter at lower redshifts, whereas a volume-limited criterion identifies a uniform sample of galaxies that occupy the same luminosity interval at different redshifts.
In addition, the cross- and auto-correlation calculations are evaluated at different 
redshift in a flux-limited sample. The redshift number density of LRGs peaks at $z\sim 0.45$, while the Mg\,II redshift distribution is flat. 
This implies that the mean redshifts of the cross-correlation and auto-correlation calculations are different, introducing additional uncertainties in the estimated mean halo bias. This problem is alleviated in a 
volume-limited sample because the cross- and auto-correlation terms have 
similar redshift evolution (see Figure 1).  Comparing the correlation functions determined using different subsamples allowed us to evaluate possible systematic uncertainties due to sample selections.

\begin{figure}
 \vspace{5pt}  \centerline{\hbox{ \hspace{0.0in}
     \includegraphics[angle=-90,scale=0.40]{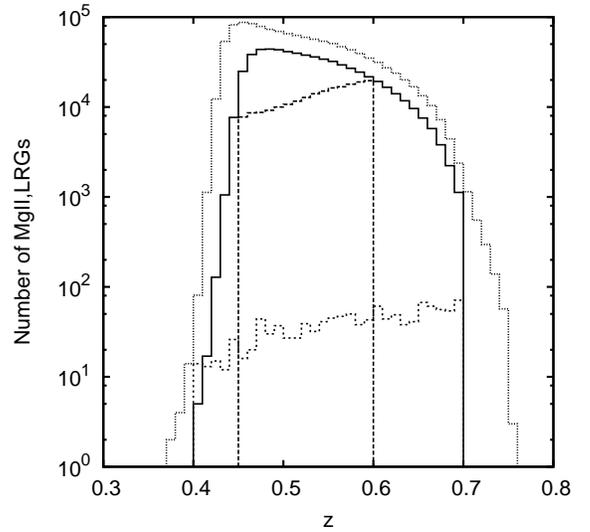}
      }
   } \vspace{5pt}
   \caption{ Redshift distributions of the initial MegaZ LRG sample
     and of the three LRG samples used in our correlation studies.
     The initial MegaZ catalog is shown in the \emph{dotted} line. The
     \emph{solid} line shows the flux-limited sample ($z=0.40-0.70$)
     and the \emph{dashed} line shows the volume-limited sample
     ($z=0.45-0.6$).  The redshift distribution of Mg\,II absorbers in
     our primary sample is overplotted in \emph{short-dashed} line for
     comparison.  }
\label{lrgs_z_catalog}
\end{figure}

\begin{figure}
  \vspace{0.5pt}
\centerline{\hbox{ \hspace{0.0in}
     \includegraphics[angle=-90,scale=0.40]{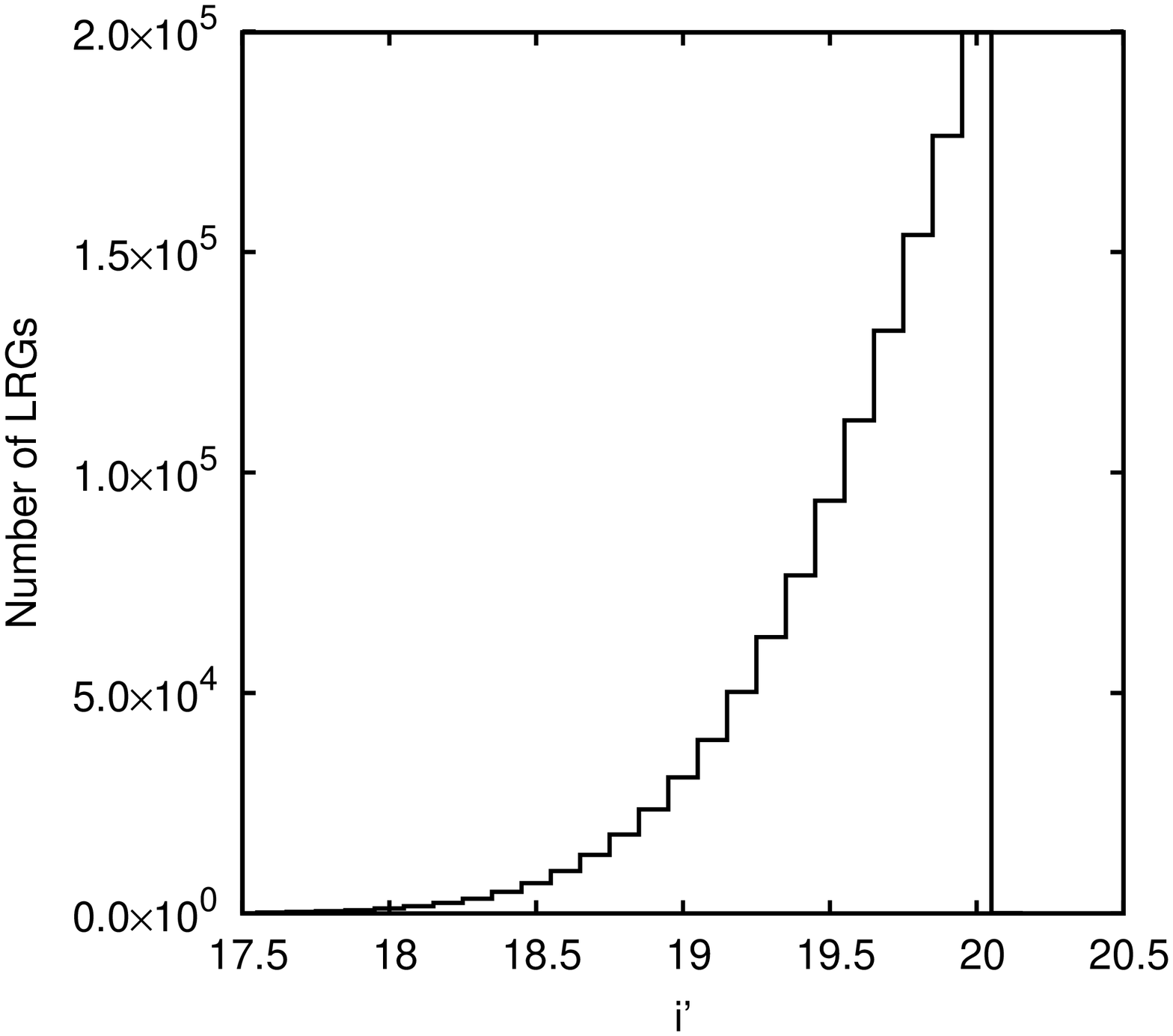}
    }
   }
 \vspace{0.5pt}
  \vspace{5pt}
  \centerline{\hbox{ \hspace{0.0in}
       \includegraphics[angle=-90,scale=0.40]{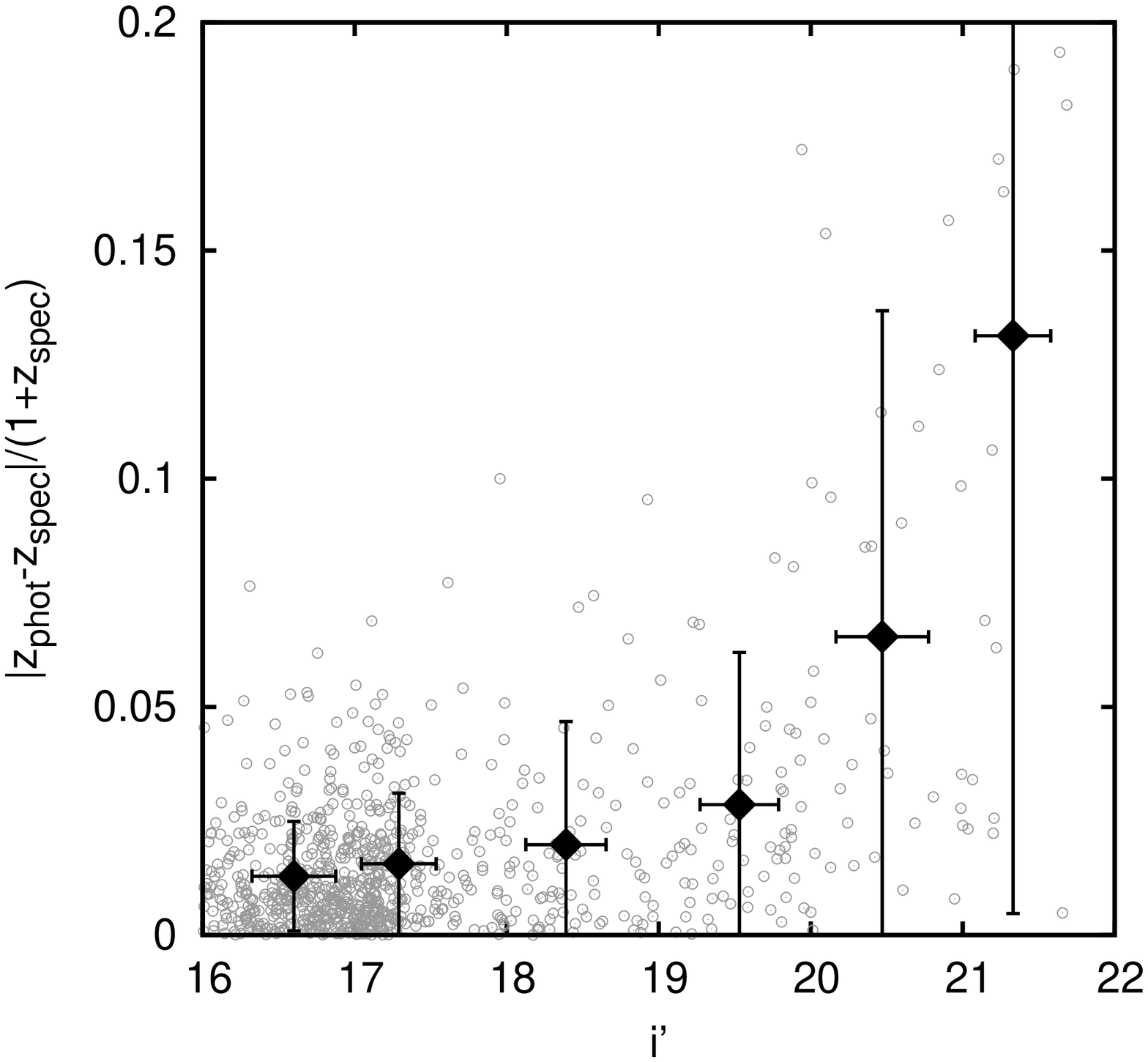}
     }
    }
\vspace{5pt}
\caption{\emph{Top panel}: 
  Magnitude distribution of the LRGs satisfying our selection
  criteria ($\approx 9.6\times10^5$ LRGs).  The line
  shows the $i'$ magnitude distribution.  Only
  galaxies with $i' < 20$ are included in our LRG sample.
  \emph{Bottom panel}: Photometric redshift error versus $i'$
  magnitude for a subsample of 1,000 galaxies in SDSS taken from
  \citet{oyaizu2008a}.  The photometric errors are listed in the SDSS
  Table "photozerr2" under "CC2".  Black diamond points with error
  bars represent the mean and dispersion calculated in bins of $\Delta
  i'=1$ mag.  }
\label{lrgs_photoz} 
\end{figure}

The limiting absolute magnitude of 
the volume-limited sample was determined by calculating the absolute magnitude of
our faintest galaxies ($i'=20.0$) at the limiting redshift $z=0.60$
\begin{equation}
20.0-{M}_i = {\rm DM}(z) + k(z) \; ,
\label{absolute_magnitude}
\end{equation}
where ${\rm DM}(z)$ is the distance modulus and $k(z)$ represents the
$k$-correction.  At $z=0.60$, the limiting magnitude is $M_i - 5\log h
= -22$ and we found roughly 197\,K galaxies more luminous than this
limit. A quantitative description of the three LRG samples including
the number of objects and redshift range can be found in Table
\ref{summary_calculations}.  Figure \ref{lrgs_z_catalog} shows the
redshift distributions of the three LRG samples and the top panel of Figure
\ref{lrgs_photoz} shows the magnitude distributions of our 
flux-limited LRG sample and of the initial LRG sample.

It is notoriously difficult to determine the photometric redshift of
faint galaxies.  The bottom panel of Figure \ref{lrgs_photoz} shows a
subsample of the photometric redshift errors taken for 1,000
overlapping galaxies between MegaZ and the SDSS photometric redshift
table. The photoz's errors shown in this figure are taken in the
"photozcc2" table found on the SDSS skyserver archive
\citep{oyaizu2008a}.  The photometric redshift errors obtained by the
MegaZ team are consistent with the ones in
\citet{oyaizu2008a} to within measurement uncertainties.
However, it is clear from Figure \ref{lrgs_photoz} 
that for $i'>20$, the uncertainties in the photoz's increase rapidly. For this reason, we decided to restrict ourselves 
to LRG candidates with $i' <20$.
This procedure yielded a total of 962,216 MegaZ objects satisfying our
additional selection criteria.  This constitutes our \emph{primary}
LRG catalog.

\section{Mg\,II absorbers-LRG cross-correlation}
\subsection{Method}
We used the \citet{landy1993a} (LS93) minimum variance estimator to calculate 
the projected two-point correlation statistics between Mg\,II absorbers and LRGs.
Using the LS93 estimator, the real-space correlation function can be calculated 
following  
\begin{equation}
\xi(r_p,\pi) = \frac{D_{\rm a}D_{\rm g} - D_{\rm a}R_{\rm g} 
- D_{\rm g}R_{\rm a} + R_{\rm a}R_{\rm g}} {R_{\rm a}R_{\rm g}}
\label{landy}
\end{equation}
where $R$ and $D$ are random points and data, the subscripts $a$ and
$g$ refer to absorbers and galaxies, $r_p$ is the projected comoving
separation between two objects on the sky and $\pi$ is their distance
parallel to the line of sight.  Note that the above two-point
estimator has been successfully used in previous correlation study
based on QSO absorbers and galaxy data collected from a smaller area
survey \citep{adelberger2003a}.  In practice, we calculated the
projected two-point correlation statistics by summing all pairs along
the sightline
\begin{equation}
w_p(r_p) = \int_l \xi(r_p,\pi) d\pi
\end{equation}
and inside of redshift limits $l$ of our three LRG samples. 

Another commonly used estimator, $1+\xi(r_p,\pi) = \frac{D_{\rm a}D_{\rm
    g}}{D_{\rm a}R_{\rm g}}$ has also been used in the initial stages
of this work for comparison and validation purposes.  Even though this
estimator is easier (no random absorbers) and faster (two terms to
compute instead of four), the variance of $w_p$ is larger for this 
estimator than for LS93. For this reason, we employed the LS93 estimator for
all $w_p$ calculations.

We divided the pairs into eight $r_p$ bins covering the range $0.2-35~h^{-1}$ Mpc. 
The bins were equally separated in logarithm space. The bin size and the $r_p$ value 
of the inner most bin were determined in such a way that at least 10 $DD$ pairs 
were found in that bin. The upper limit of $35~h^{-1}$ Mpc was chosen to be a few times 
smaller than the size of the jackknife cells (see section 3.1.3).

\subsubsection{Survey mask}
For the calculation of the two-point statistics (following equation \ref{landy}) 
both data and randoms were distributed over the exact same 
survey mask. 
Different masks for LRGs and MgII would alter the shape and amplitude of the 
correlation signal in an undesirable fashion, especially at large separations.
To make sure the sky coverage was the same for Mg\,II, LRGs and their randoms, 
we used the lowest common denominator for all : the DR4 spectroscopic sky survey mask. 
Indeed, Mg\,II absorbers were taken 
from the SDSS DR5 spectroscopic sample which includes the QSO sightlines inside the DR4 spectroscopic sky.
The LRGs were found in the DR4 imaging sky which encompasses the DR4 spectroscopic sky. The main disadvantage of using 
this mask was that the number of rejected objects falling outside of the mask was large, $\approx 40\%$.

To be able to determine which Mg\,II absorbers and LRGs fell inside
this DR4 spectroscopic mask, we used the mask catalog provided by the
NYU Value-Added Galaxy Catalog team \citet{blanton2005a}.  The mask
corresponds to the angular selection function describing the
completeness of the SDSS spectroscopic across the sky. It is defined
by spherical polygons.  The completeness quantifies the fraction,
inside each polygon, of galaxies with spectrocopic redshift.  We used 
the angular selection function
of the spectroscopic SDSS DR4 sky ($\rm{sdssdr4safe0res6d}$).  

\subsubsection{Generating random Mg\,II absorbers and LRGs}

The right ascension and declination of the LRGs were randomly selected
over the DR4 spectroscopic sky using the function \emph{ransack}
available in the Mangle software package
\citep{hamilton2004a,swanson2008a}. The redshifts of the random
galaxies were determined by sampling the redshift distribution of the
LRG dataset.  Determining the sky positions and redshifts of the
random absorbers was not a straightforward process.  The angular
selection function of quasars follows the DR4 spectroscopic mask, but
the mask defining the positions where absorbers can be found is
limited to the actual coordinates of the QSOs themselves.  Thus, the
random absorbers must be distributed randomly among fixed QSO
sightlines for which SDSS spectra are available.  Assigning random
Mg\,II this way eliminates any undesired bias due to the intrinsic
clustering of QSO sightlines. We identified these sightlines
from the \citet{schneider2007a} SDSS DR5 QSO catalog by selecting all
QSOs falling inside our DR4 spectroscopic mask with redshifts large
enough to allow for the detection of Mg\,II absorbers inside the
redshift range of interest for our calculations ($z=0.40-0.70$). We
found a total of $\approx 5.5~\times~10^4$ sightlines. The (ra,dec)
positions of the random absorbers were determined by randomly selecting
the coordinates of these sightlines. Redshifts were selected randomly
from a top hat probability distribution function over $z=0.4-0.7$.

The number of random LRGs, $R_g$, and the number of random Mg\,II absorbers, $R_a$, were
determined after running convergence tests with increasing number of
randoms. We varied the number of each random set (Mg\,II and LRGs)
independently until the measured $w_p$ approached asymptotic values.  The numbers of randoms
used in our $w_p$ calculation were $R_a = 10^5$ and $R_g =
4\times10^6$ for Mg\,II and LRGs, respectively.  It is important to
note that increasing the number of random Mg\,II absorbers to a very
large number in comparison with the number of available QSO sightlines
($\sim$ 55K) would not lead to a rapid convergence because we would
simply be overcounting the same $D_{\rm g}R_{\rm a}$ pairs and no
additional information would be gained.

\subsubsection{Relative contribution of cosmic variance and 
photometric uncertainty to $w_p$ errors}

We looked at two independent sources contributing to the error bars on
$w_p$ : cosmic variance and photometric redshift errors.  Photometric
redshift errors are expected to affect the measurements of $w_p$'s in
two different ways.  First, redshift uncertainties are expected to
increase the noise in the $w_p$ measurements due to uncertainties in
the object positions.  Second, redshift uncertainties are expected to
systematically alter the $w_p$ measurements to lower values due to an
inherent sample selection bias.  A galaxy sample selected based on
imprecise photometric redshifts contain galaxies from a broader
redshift range than one selected based on precise spectroscopic
redshifts, and consequently reducing the observed correlation signal
by including additional uncorrelated pairs.  In this section, we
address random errors in the $w_p$ measurements due to cosmic variance
and object distance uncertainties.  We defer the discussion on the
systematic errors of $w_p$ due to the sample selection bias to \S\
3.2.2.

We estimated the cosmic variance using the jackknife resampling
technique.  The sky was separated into $N=192$ (see section 3.2.1 
for a justification of this number) cells of roughly equal
survey area.
The cosmic variance for each point $r_{p,i}$
corresponds to the $i$th-diagonal element of the covariance matrix was
calculated using a jackknife resampling technique :
\begin{equation}
{\rm COV}(w_i,w_j)=\frac{N-1}{N}\sum_{k=1}^{N}(w_i^k-\overline{w_i})(w_j^k-\overline{w_j})
\label{covariance}
\end{equation}
and $k$ represents the iteration in which box $k$ was removed. 
The mean $\overline{w_{i}}$ was calculated for bin $i$ over all $w_{p}^k$'s. 

The impact of the large uncertainties due to photometric redshifts on
the size of the $w_p$ errors is not taken into account by the
jackknife resampling technique alone.  Previous works (e.g.,\
\citealt{bouche2006a,blake2008a}) have focused primarily on cosmic
variance in the calculation of the errors bars, but have not addressed
additional random noise due to photometric redshift errors which are
quite large  ($\sigma_z \approx 0.05$ for $i'<20$) compared to the 
redshift ($z=0.5$).

To account for the independent contribution of photoz's uncertainties
on the final $w_p$ error bars, we generated 100 independent
realizations of the MegaZ catalog.  For each one of them, we resampled
the redshift of each individual galaxy according to a normal
distribution $N(z_{\rm phot},\sigma_{z})$, where $z_{\rm phot}$ is the
photometric redshift and $\sigma_z$ is the photometric redshift error
of each galaxy. In this case, we followed the error function
$\sigma_z=0.03(1+z)$ found in \citet{collister2007a} to assign photometric redshift
errors to galaxies.  A new mock LRG sample was then established
according to the criteria discussed in \S\ 2.2.  We calculated $w_p$
for each one of these realizations and assigned the error contribution
from redshift uncertainties to the final error budget of $w_p$ by
calculating the dispersion among these 100 independent realizations.

We found that the size of the $w_p$ error bars was dominated by cosmic
variance. The contribution of photometric redshift errors to the $w_p$
error budget was small, at most 20\% for the smallest separation
bin. It is negligible ($\sim 1\%$) at large separations where we
calculated the relative clustering strength and absolute bias.  For
this reason, we decided to adopt the cosmic variance $\sigma_{i}$ as
the error on $w_p$ and neglect the contribution of the photometric
redshifts.  

\subsection{Results}

This section addresses the cross- and auto-correlation results for the
three LRG samples and the effect of photometric redshifts on the
clustering amplitude. For each one of our three LRG samples, we
considered three subsamples of Mg\,II absorbers:
weak $W_r(2796)=1-1.5$ \AA, strong $W_r(2796)=1.5-5$ \AA, and all
absorbers.  A description of each correlation calculation can be found
in Table \ref{summary_calculations}. We also include the number of
$DD$ pairs (equation \ref{landy}) in the first bin and the number of
data found in each subsample in columns (6), (7), and (8).

\begin{centering}
\begin{deluxetable*}{lccccccc}
\tabletypesize{\scriptsize}
\tablecaption{Summary of the cross- and auto-correlation calculations}
\tablewidth{0pt}
\tablehead{
&  &  & \colhead{$W_{\rm{r,min}}(2796)^a$} &
\colhead{$W_{\rm{r,max}}(2796)$} & \colhead{Number} & \colhead{Number} & \colhead{Number} \\
\colhead{Sample} & \colhead{$z_{\rm min}$} & \colhead{$z_{\rm max}$} & \colhead{(\AA)} &
\colhead{\AA} & \colhead{DD pairs$_{\rm (1st bin)}$$^b$} & \colhead{LRGs} & \colhead{Mg\,II} \\
\colhead{(1)} & \colhead{(2)} & \colhead{(3)} & \colhead{(4)} & \colhead{(5)} & \colhead{(6)} & \colhead{(7)} & \colhead{(8)} 
}
\startdata
\multicolumn{8}{c}{Volume-limited sample} \\
\cline{1-8}\\
V-weak(VW) & 0.45 & 0.60 & 1.0 & 1.5 & 21 & 197,968  & 279\\
V-strong(VS) & 0.45 & 0.60 & 1.5 & 5.0 & 13 & 197,968 & 257  \\
V-all(VA) & 0.45 & 0.60 & 1.0 & 5.0 & 34 & 197,968 & 536  \\
\cutinhead{Flux-limited samples}
F1-weak(F1W) & 0.45 & 0.60 & 1.0 & 1.5 & 47 & 517,549 & 279 \\
F1-strong(F1s) & 0.45 & 0.60 & 1.5 & 5.0 & 35 & 517,549 & 257 \\
F1-all(F1A) & 0.45 & 0.60 & 1.0 & 5.0 & 82 & 517,549 & 536 \\
F2-weak(F2W) & 0.40 & 0.70 & 1.0 & 1.5 & 88 & 618,086 & 541 \\
F2-strong(F2S) & 0.40 & 0.70 & 1.5 & 5.0 & 68 & 618,086 & 617 \\
F2-all(F2A) & 0.40 & 0.70 & 1.0 & 5.0 & 156 & 618,086 & 1158 \\
\cutinhead{LRGs auto-correlation}
V-LRGs(VG) & 0.45 & 0.60 & - & - & 6,977 & 197,968 & - \\
F1-LRGs(F1G) & 0.45 & 0.60 & - & - & 43,093 & 517,549 & - \\
F2-LRGs(F2G) & 0.40 & 0.70 & - & - & 55,062 & 618,086 & - \\
\enddata
\tablenotetext{a}{$W_r$ is the rest-frame equivalent width. }
\tablenotetext{b}{The first bin is centered on $0.312~h^{-1}$ Mpc}
\label{summary_calculations}
\end{deluxetable*}
\end{centering}

\renewcommand{\arraystretch}{1.5}

\subsubsection{Cross- and auto-correlation results}

\begin{figure*}
\centerline{
\includegraphics[angle=-90,scale=0.90]{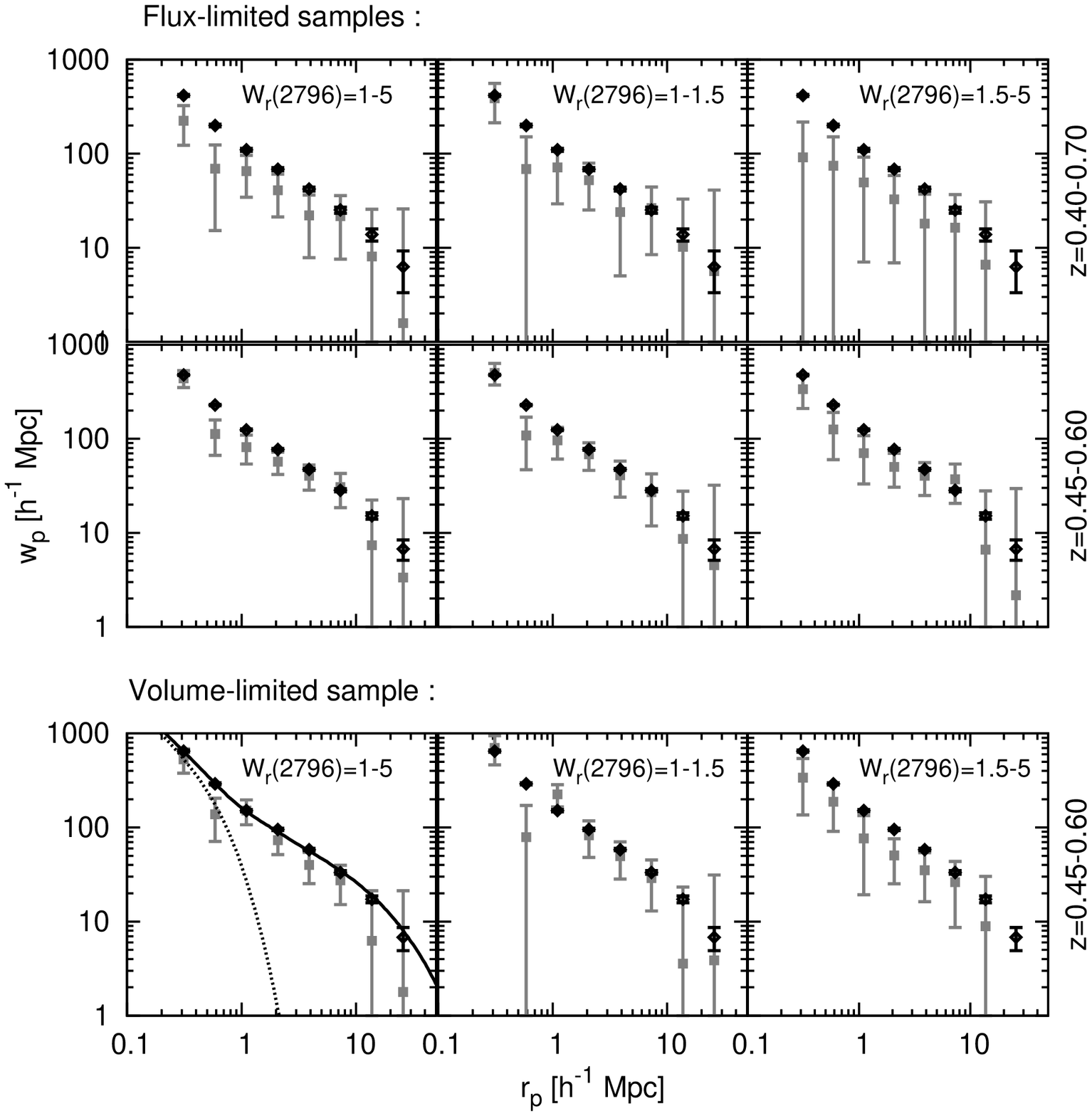}
}
\caption{ Cross- and auto-correlation results for the flux and
  volume-limited samples of LRGs. From \emph{top to bottom},
  flux-limited sample $z=0.40-0.70$, flux-limited $z=0.40-0.60$, and
  volume-limited $z=0.40-0.60$. From left to right, \emph{all}
  absorbers, \emph{weak} absorbers and \emph{strong} absorbers. The
  rest-frame equivalent-width coverage is listed at the top of each
  column. In all panels, the gray points represent the
  cross-correlation results (Mg\,II and LRGs) and the black points are
  for the LRGs auto-correlation. In the bottom left panel, we show the 
  halo occupation distribution analysis done on the volume-limited sample  
  of LRGs. The solid line shows the full HOD model and the dotted line 
  the one-halo contribution (see section 4.2). Note that both cross- and 
  auto-correlation results were corrected for photometric redshifts (see section 3.2.2). 
  Error bars represent the 1-$\sigma$
  jackknife errors calculated on $N=192$ sky boxes of equal area. Note
  that the last cross-correlation points in two of the strong Mg\,II
  absorber sample are negative and not displayed in these panels. The
  weak absorbers are essentially unbiased with respect to the LRGs and
  their clustering is strong, even down to 0.3 $ h^{-1}$ Mpc.  }
\label{everything}
\end{figure*}

Figure \ref{everything} shows the cross- and auto-clustering results
for the flux- and volume-limited samples of LRGs. For each one of
these samples, the results are shown for the three Mg\,II subsamples
in gray. We overplot the LRGs auto-correlation results in black.  A
clear feature of Figure \ref{everything} is that, for the three LRG
samples considered, the clustering amplitude of weak absorbers is
systematically higher than for strong ones.  It is also interesting to
note that after accounting for systematic bias due to redshift
uncertainties (see \S\ 3.2.2), weak Mg\,II absorbers appear to be
unbiased (sharing the same clustering amplitude) with respect to LRGs
in both flux-limited and volume-limited samples.

\begin{figure*}
\centerline{
\includegraphics[angle=-90,scale=0.70]{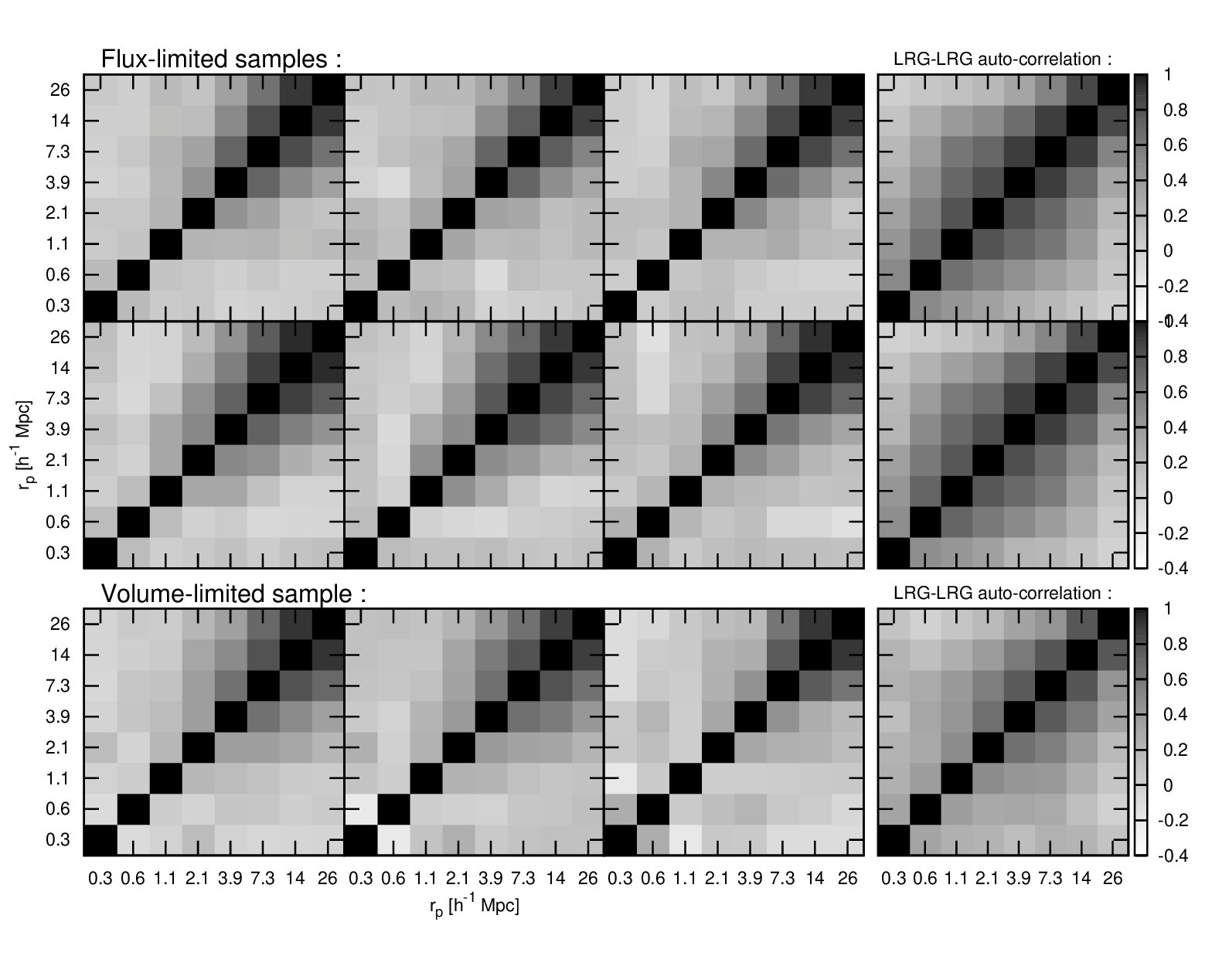}
}
\caption{Gray-scale plot of the correlation matrix (see equation
  \ref{correlation}) showing the degree of covariance between
  different separation bins. The different matrices shown correspond
  to the same calculations of Figure \ref{everything} for
  $W_r(2796)=1-5$ \AA\ (left), $W_r(2795)=1-1.5$ \AA\ (middle-left),
  and $W_r(2796)=1.5-5$ \AA\ (middle-right), respectively. The fourth
  column shows the LRGs auto-correlation results.}
\label{cij}
\end{figure*}

For each auto- and cross-correlation calculations, we estimated the
correlated uncertainties between different $r_p$'s using the
covariance matrix (see equation \ref{covariance}) and the
normalized correlation matrix
\begin{equation}
\rho(w_i,w_j)=\frac{{\rm COV}(w_i,w_j)}{\sqrt{{\rm COV}(w_i,w_i){\rm COV}(w_j,w_j)}} \; . 
\label{correlation}
\end{equation}
We performed a convergence test on the number of jackknife boxes. We
made sure that the off-diagonal elements of the correlation matrix
varied by less than 10\% after doubling the number of boxes.  We plot
in Figure \ref{cij} the normalized correlation matrix $\rho(w_i,w_j)$
for all calculations, including the LRGs auto-correlation shown in the
fourth column. Adjacent bins are strongly correlated for the
auto-correlation functions of LRGs and, in all cases of the
LRG--Mg\,II cross-correlation measurements, bins with large $r_p$ are
more correlated.

\subsubsection{Effect of photometric redshifts on the amplitude of clustering}
An important effect of photometric redshift errors is the sample
selection uncertainty.  Photometric redshift uncertainties not only
affect the precision of $w_p$, but also lower its accuracy that one
would normally get from using spectroscopic redshifts (e.g.,\
\citealt{brown2008a}).  This needs to be accounted for in the bias
calculation.  One can think of photometric redshift errors as adding
uncorrelated galaxies in the calculation, and the overall effect is an
unwanted widening of the galaxy redshift distribution.
This effect results in a systematic error in the clustering signal and the 
amplitude of $w_p$ is comparatively smaller than expected in
calculations using only spectroscopic redshifts.

To estimate the attenuation of the clustering signal due to
photometric errors, we performed several tests on mock LRG galaxy
distributions. Our mock catalog is produced by populating the halos in
an N-body simulation with a halo occupation function determined from
the spectroscopic LRG sample in \citet{zheng2008a} (specifically, the
``faint'' sample). The simulation is large, 1 Gpc/h on a side, and
represents our fiducial cosmology at $z\sim 0.5$ (see
\citealt{tinker2008b} for more details on this simulation).  Since the
box size is not big enough to cover the redshift range of our study,
we mirror-imaged three identical copies of the box along the redshift
direction.  The mock LRG catalog has precise positions and therefore
mimic a spectroscopic sample.

\begin{figure}
 \vspace{5pt}  
   \centerline{\hbox{ \hspace{0.0in}
    \includegraphics[angle=-90,scale=0.40]{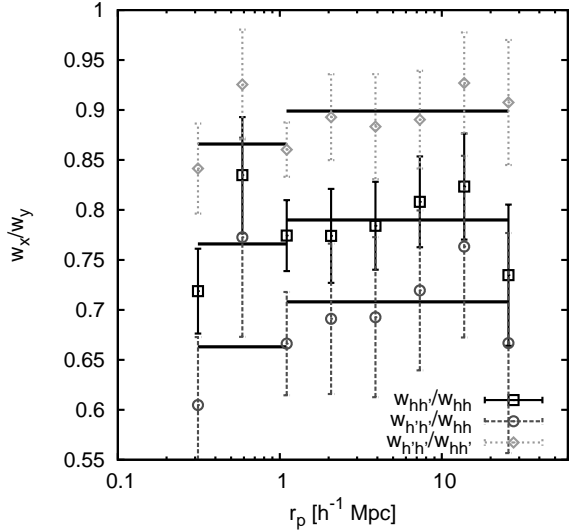}
      }
     }
 \vspace{5pt}
 \caption{Ratios of cross-correlation signals obtained from the
   simulation boxes to address the impact of photometric redshift
   uncertainties on the amplitude of $w_p$.  \emph{Solid black} :
   ratios of $w_p$'s where one of the two datasets has redshift
   position perturbed (within its error bars) ($w_{hh'}$) over $w_p$'s
   obtained where both datasets have spectroscopic redshift
   ($w_{hh}$); \emph{dashed gray} : ratios where both datasets have
   perturbed redshifts ($w_{h'h'}$) over two datasets with unperturbed
   redshifts ($w_{hh}$); \emph{Dotted light gray} : $w_p$ for
   perturbed-perturbed datasets ($w_{h'h'}$) over
   perturbed-unperturbed ($w_{hh'}$).  The last set of points gives us
   an estimate of the systematic bias in the observed clustering
   amplitude of Mg\,II absorbers using photometric redshift identified
   LRGs.  Error bars show dispersion among 100 realizations of the
   perturbed datasets ($h'$). The thick solid lines represent best-fit
   amplitude obtained for the one ($r_p \leq 1.1 h^{-1}$ Mpc) and
   two-halo ($r_p>1.1 h^{-1}$ Mpc) terms. Best-fit values are listed
   in table \ref{summary_fit_mgii_perturbation}. The best-fit
   amplitude obtained for the light gray points is used to correct for
   the addtional systematic error introduced by using photoz's in the
   calculation of the correlation functions.  }
\label{simbox}
\end{figure}

Three tests were performed. First, we auto-correlated the mock LRG
sample without applying any modifications to their redshifts. This
first case mimics the results one would get by auto-correlating a
sample of LRGs with spectroscopic redshifts (the $w_{hh}$ case in Figure 6). 
Second, we cross-correlated the mock LRG sample with one
that involved perturbed redshifts.  The perturbed redshifts were
generated by sampling a normal distribution centered on the input
redshift in the mock catalog within $\sigma_z=0.03(1+z)$.  That is
the $w_{hh'}$ term in Figure \ref{simbox}, which is in analogous to our
cross-correlation measurements betwen absorbers with spectroscopic
redshifts and galaxies with photometric redshifts.  Third, we
auto-correlated the mock LRG sample with perturbed redshifts
($w_{h'h'}$ in Figure 6). This mimics our LRG auto-correlation function.

We then computed the $w_p$ ratios between the different calculations
and plotted the results in Figure \ref{simbox}.  The three sets of
symbols correspond to ratios of $w_p$ calculated over the same series
of $r_p$ bins as in the correlation calculations shown in Figure
\ref{everything}.  We show, by thick black lines, the best-fit
amplitude of the three ratios in two different regimes : small and
large separations.  Best-fit values are listed in Table
\ref{summary_fit_mgii_perturbation} with their corresponding
uncertainties ($\Delta \chi^2 < 1$). Error bars on the ratios
themselves correspond to the dispersion obtained among 100
realizations of a mock catalog with perturbed redshifts.  The best-fit
amplitude (of the light points) at large separation enters in the
calculation of the relative bias (see section 5.1).

The dark datapoints in Figure \ref{simbox} show that the
cross-correlation function between Mg\,II absorbers and photometric
redshift identified LRGs is underestimated by roughly 20\% due to
redshift errors.  The gray points show that the auto-correlation
function of these LRGs is underestimated by roughly 30\%.  Finally,
the light-gray points show that the mean bias calculated based on the
ratio of absorber-LRG cross-correlation and LRG auto-correlation
functions is {\it overestimated} by roughly 10\%.  

Our estimated reduction in clustering strength of 10\% is smaller than
the value quoted in \citet{bouche2004a}. These authors found that, in
the case of a gaussian redshift distribution for galaxies, the
amplitude of the MgII-LRG cross-correlation is overestimated by 25
$\pm$ 10\%. They used numerical integration and mock galaxy catalog
with phometric redshift uncertainties $\Delta_z=0.1$ corresponding to
size of their redshift interval of interest. The discrepancy could be
partly attributed to the larger photometric redshift uncertainties
used by these authors.

\begin{centering}
\begin{deluxetable*}{lcccc}
\tabletypesize{\scriptsize}
\tablecaption{Estimated reduction factor of the cross-correlation amplitude due to photometric 
redshift sampling errors}
\tablewidth{0pt}
\tablehead{ \colhead{Measurement} & \colhead{$r_p < 1~h^{-1}$ Mpc} & \colhead{$\chi^2$/D.O.F} & \colhead{$r_p>1~h^{-1}$ Mpc } 
& \colhead{$\chi^2$/D.O.F}}
\startdata
$w_{\rm Mg\,II-LRG}$ & 0.77 $\pm 0.03$ & 1.35 & 0.79 $\pm 0.02$ & 0.26 \\
$w_{\rm LRG-LRG}$ & 0.66 $\pm 0.04$ & 0.97 & 0.71 $\pm 0.04$ & 0.12 \\
\enddata
\label{summary_fit_mgii_perturbation}
\end{deluxetable*}
\end{centering}

\begin{figure*}
 \vspace{5pt}
   \centerline{\hbox{ \hspace{0.0in}
    \includegraphics[angle=0,scale=0.9]{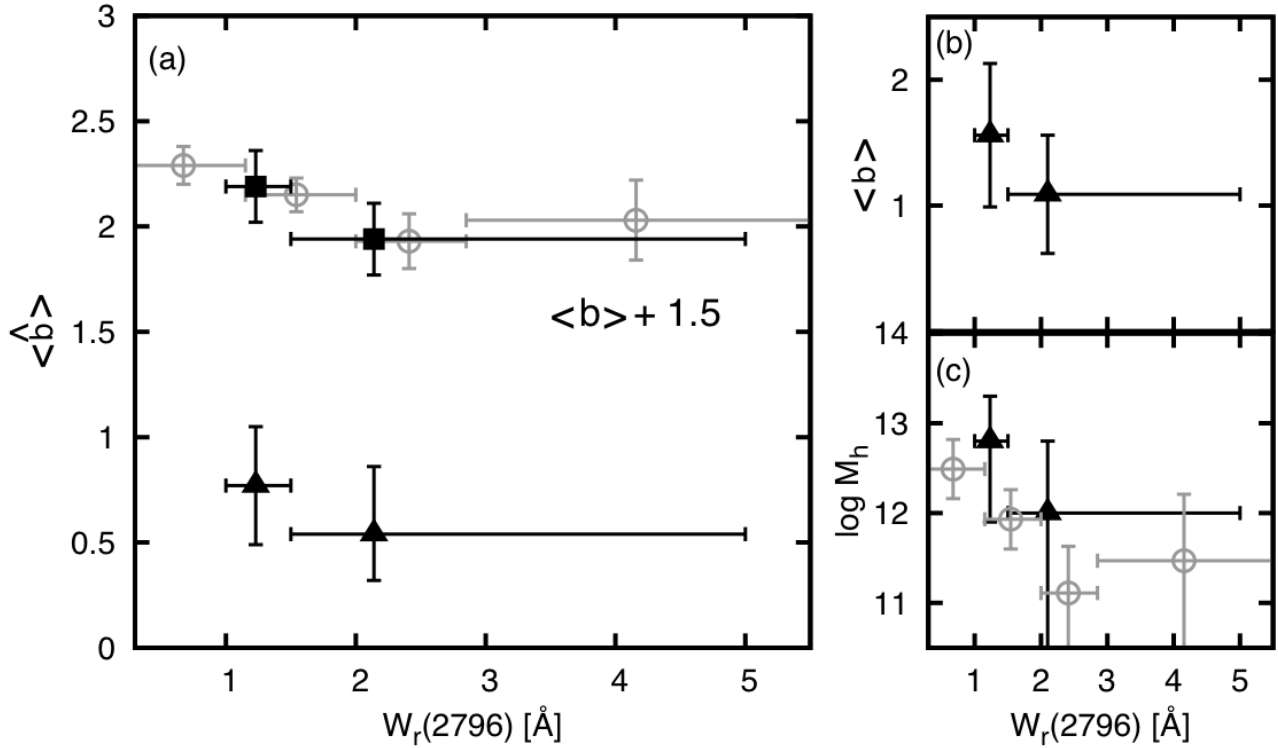} 
     }
     }
 \vspace{5pt}
\caption{\emph{Panel (a)} shows the relative bias of Mg\,II for two of the three LRG samples. 
\emph{Solid lines with triangles} are for the volume-limited sample, \emph{solid lines with squares} for the 
flux-limited $z=0.40-0.70$, and \emph{open circles} are taken from \citet{bouche2006a} for direct comparison 
with the flux-limited sample. For each LRG sample, the points are derived 
from the direct ratio of the points for $r_p > 1~h^{-1}$ Mpc. The power-law technique yielded very similar results.  
We shifted the results for the flux-limited samples 
by $\langle \hat{b} \rangle+1.5$ for more clarity. 
\emph{Panel (b)} : absolute bias for the volume-limited sample. 
\emph{Panel (c)} bias-inverted halo mass derived from the absolute bias. The bias-weighted halo mass results of \citet{bouche2006a} are shown
in open circles. Error bars along the x-axis represent the binning used. }
\label{bias}
\end{figure*}

\begin{centering}
\begin{deluxetable*}{lcccc}
\tabletypesize{\scriptsize}
\tablecaption{Best-fit power-law $f(x)=ax^b$ results for the cross and auto-correlations.}
\tablewidth{0pt}
\tablehead{
\colhead{Sample} & \colhead{$a$} & \colhead{$b^{\rm i}$} & \colhead{$\chi^2$} & \colhead{$\chi^2/{\rm D.O.F}^{\rm ii}$}
}
\arraystretch
\startdata
VW & 0.337 $\pm 0.11 $ & -0.835  & 1.2  & 0.3  \\
VS & 0.206 $\pm 0.09 $ & -0.835 & 0.394 & 0.13  \\
VA & 0.275 $\pm 0.075 $ & -0.835 & 1.06  & 0.27 \\
F1W & 0.245 $\pm 0.080$ & -0.783 & 0.54 & 0.14 \\
F1S & 0.194 $\pm 0.070$ & -0.783 & 4.60 & 1.15 \\
F1A & 0.206 $\pm 0.056$ & -0.783 & 3.78 & 0.76 \\
F2W & 0.160 $\pm 0.046$ & -0.781 & 3.47 & 0.87 \\
F2S & 0.114 $\pm 0.047$ & -0.781 & 0.776 & 0.258\\
F2A & 0.136 $\pm 0.035$ & -0.781 & 3.68 & 0.92 \\
\cutinhead{LRGs auto-correlation}
VG & 0.356 $\pm 0.013$ & -0.835 $\pm 0.029$ & 11.25 & 3.75 \\
FG1 & 0.273 $\pm 0.008$ & -0.783 $\pm 0.026$ & 9.72 & 3.24  \\
FG2 & 0.241 $\pm 0.007$ & -0.781 $\pm 0.026$ & 6.79 & 2.26  \\
\enddata
\tablenotetext{i}{In the case of the LRGs auto-correlation calculation, both $a$ and $b$ are
free parameters. For cross-correlation Mg\,II-LRGs, $b$ is kept fixed and corresponds to the best-fit value
obtained for the auto-correlation.}
\tablenotetext{ii}{$\chi^2$ per degree of freedom. For auto-correlation calculations, D.O.F=4 and D.O.F=5 for cross-correlation.}
\label{power_law}
\end{deluxetable*}
\end{centering}

\begin{centering}
\begin{deluxetable*}{lcccccc}
\tabletypesize{\scriptsize}
\tablecaption{Relative and absolute bias. Bias-inferred masses.}
\tablewidth{0pt}
\tablehead{
\colhead{} & \colhead{direct} & \colhead{power-law} & \colhead{direct} & \colhead{power-law} & \colhead{direct} & \colhead{power-law} \\ 
\colhead{Sample} & \colhead{$\langle \widehat{b}_{\rm{DR}}^\dagger \rangle$} & \colhead{$\langle \widehat{b}_{\rm{RA}} \rangle $} 
                 & \colhead{$\langle b_{\rm{DR}}^{\ddagger} \rangle $} & \colhead{$\langle b_{\rm{RA}} \rangle$} 
                 & \colhead{$(\log M_{\rm{h}})_{\rm{DR}}$} 
                 & \colhead{$(\log M_{\rm{h}})_{\rm{RA}}$}\\
\colhead{(1)} & \colhead{(2)} & \colhead{(3)} & \colhead{(4)} & \colhead{(5)} & \colhead{(6)} & \colhead{(7)}
}
\arraystretch
\startdata
\colhead{} & \multicolumn{2}{c}{Relative bias} & \multicolumn{2}{c}{Absolute bias} & \multicolumn{2}{c}{Mean mass$^{i}$}\\
\cline{2-3}  \cline{4-5}  \cline{6-7} \\
VW & 0.77  $\pm 0.28 $ & 0.85 $\pm 0.25$  & 1.56 $\pm 0.57 $ & 1.72  $\pm 0.51 $ & $12.8^{+0.5}_{-1.1} $ & $13.0^{+0.4}_{-0.6} $  \\
VS & 0.54  $\pm 0.23 $ & 0.52 $\pm 0.21$  & 1.09 $\pm 0.47 $ & 1.05  $\pm 0.42 $  & $12.0^{+0.8} $  & $11.9^{+0.8} $    \\
VA & 0.67  $\pm 0.19 $ & 0.69 $\pm 0.16$  & 1.36 $\pm 0.38 $ & 1.41  $\pm 0.33 $ & $12.5^{+0.4}_{-0.8} $  & $12.6^{+0.4}_{-0.6} $ \\
F1W &0.80 $\pm 0.26 $ & 0.81 $\pm 0.24$  & -  & - & - & - \\
F1S &0.75 $\pm 0.25 $ & 0.64 $\pm 0.21$ & - & - & - & - \\
F1A &0.74 $\pm 0.19 $ & 0.68 $\pm 0.16$ & - & - & - & - \\
F2W &0.69 $\pm 0.17 $ & 0.60 $\pm 0.15$ & - & - & - & -  \\
F2S &0.44 $\pm 0.17 $ & 0.43 $\pm 0.16$ & - & - & - & - \\
F2A &0.56 $\pm 0.13 $ & 0.51 $\pm 0.12$ & - & - & - & - \\
\enddata
\tablenotetext{$\dagger$}{The subscript DR represents the results obtained with the mean ratio of the two-halo term data points and RA is calculated 
from the ratio of the best-fit power-law amplitudes of the cross- and auto-correlations.}
\tablenotetext{$\ddagger$}{$b$ is the absolute bias.}
\tablenotetext{$i$}{Note that we did not quote a lower limit on the halo mass for strong absorbers. Indeed, the bias vs. halo 
mass relationship flattens to $b_h(M) \sim 0.7$ for $\log M_h < 9$. 
When the lower limit on the bias is $<$ 0.7, this gives us no lower limit on the halo mass.}
\label{bias_table}
\end{deluxetable*}
\end{centering}

\section{Determining the Absolute Bias and Mean Mass Scale of Absorber hosts}

\subsection{Theoretical Framework}

The bias of dark matter halos can be defined as the ratio between the
clustering of halos (at a fixed mass) and the underlying clustering of
dark matter, 

\begin{equation}
b_h^2(M,r) = \frac{\xi_h(M,r)}{\xi_m(r)}
\end{equation}

\noindent where $\xi_m(r)$ is the correlation function of the dark
matter itself. At large scales, linear bias holds and $b_h$ is
independent of $r$. In the translinear regime, $r\lesssim 10$ $h^{-1}$ Mpc,
$b_h$ has a scale dependence (with respect to either $\xi_m(r)$
obtained from linear theory or the true non-linear
clustering). Although the scale dependence of halo bias varies with halo mass,
over the mass range 
probed by LRGs and Mg\,II absorbers, $\sim 10^{12-13}$ \hmsol, the scale
dependence is nearly independent of mass and will divide out in the
cross-correlation function \citep{tinker2009a}. The auto-correlation function
of LRGs can then be expressed as

\begin{equation}
\xi_g(r) = b^2_h(M_g,r)\xi_m(r) = b^2_g(M_g)f_g^2(r)\xi_m(r)
\end{equation}

\noindent where $M_g$ is the bias-weighted mean mass scale (see
equation \ref{bw_mass}) of galaxies, $b_h$ and $b_g$ are the large-scale linear
biases of dark matter halos and galaxies, and $f(r)$ is the
scale-dependent bias term (see, e.g.,
\citealt{tinker2005a,tinker2009a}). The cross-correlation is then

\begin{equation}
\xi_{ga}(r) = b_g(M_g)b_a(M_a)f_g(r)f_a(r)\xi_m(r).
\end{equation}

\noindent Thus the relative bias of absorbers to LRGs is the ratio of
the cross to auto-correlation functions, ie, 

\begin{equation}
\hat{b} \equiv \frac{b_a}{b_g}=\frac{w_{ag}}{w_{gg}},
\label{relative_bias}
\end{equation}

\noindent which should be close to a constant at large separations.
We used the measured $w_{gg}(r_p)$ to obtain $b_g$, which we utilized
to obtain $b_a$. At $z=0.5$ and at masses above $M\sim 10^{11.5}$
\hmsol, the bias of dark matter halos increases monotonically with
mass, thus the equivalent dark matter halo mass can be obtained from
inverting the $b_h(M)$ formula. We used the following halo bias
function \citep{tinker2009a} for halos defined at an overdensity of
200 times the background

\begin{equation} 
b(\sigma)=1-A\frac{\sigma^{-a}}{\sigma^{-1}+1} + B\sigma^{-b}+C\sigma^{-c}
\label{halo_bias}
\end{equation}

\noindent where $\sigma$ is the linear matter variance on the
Lagrangian scale (radius of the halo in the initial 
mass distribution when $\delta(\rho) = (\rho-\bar{\rho})/\bar{\rho}\sim 0$) 
of the halo, $R=(3M/4\pi\bar{\rho})^{1/3}$,
and $a$, $b$, $c$, $A$, $B$, $C$ are constants ($a=0.132$, $b=1.5$,
$c=2.4$, $A=1.04$, $B=0.4$, $C=0.99$).

\subsection{The bias of LRGs}

We obtained the bias of LRGs through halo occupation modeling of the
$w_{gg}$ data and the number density of galaxies in the sample. Note
that $w_{gg}$ was corrected for the systematic error due to
photometric redshift uncertainties (see \S\ 3.2.2). Our modeling was
similar to the one performed in \citet{zheng2008a}, which is based on
the analytic halo occupation model developed in \citet{zheng2004a} and
\citet{tinker2005a}. The best-fit model is shown in the bottom left
panel of Figure \ref{everything}, which yielded a $\chi^2$ of 9.8
using the full covariance matrix.  Halo occupation models separate
pairs from galaxies located inside the same dark matter halo (one-halo
term) and pairs from galaxies located in two different halos (two-halo
term). The one-halo contribution to the clustering amplitude of LRGs
is shown in Figure \ref{everything}. At small separations, ($r_p
\lesssim 1~h^{-1}$ Mpc) the one-halo term dominates but the two-halo
term shapes the clustering signal for $r_p \gtrsim 1~h^{-1}$ Mpc.
Because the analytic model fully incorporates the scale-dependent bias
of dark matter halos, we can obtain the linear bias directly from
these data. The bias of LRGs in our sample is $b_g=2.023\pm
0.006$. The high precision is due to the large volume of the sample,
yielding an excellent measurement of the amplitude of $w_{gg}$ in the
two-halo regime.

\subsection{The relative bias of absorbers}

We calculated the relative bias using equation (\ref{relative_bias})
on large scales only ($>1.0\, h^{-1}$ Mpc). Two methods were employed.
In the first case, we fitted the cross- and auto-correlation results
using a power-law model and estimated the relative bias using the
ratio of the best-fit amplitudes.  This is a standard procedure that
has been commonly done in previous works (e.g.,\
\citealt{davis1983a}). However, the power-law model does not have a
physical justification (e.g.,\ \citealt{blake2008a,zehavi2004a}). It
simply provides an adequate fit to the data. In the second case, we
directly calculated the relative bias by taking a weighted mean ratio
of all points at $r_p>1~h^{-1}$ Mpc.  Because the measurements and
measurement errors vary significantly between data points at different
$r_p$'s, we adopted the weights $\omega_i$'s that were designed to
maximize the significance of the mean relative bias
$\langle\hat{b}\rangle$,
\begin{equation}
\langle \hat{b} \rangle = \sum_{i=4}^8 \omega_i \frac{w_{ag,i}}{w_{gg,i}}
\label{dr}
\end{equation}
where 
\begin{equation}
\omega_i=\frac{w_{ag,i}}{\sigma_i^2 w_{gg,i}}  \;. 
\end{equation}
the index $i$ denotes the $r_p$ bin and $\sigma_i$ is the associated error of
$w_{ag,i}/w_{gg,i}$ computed using the error propagation
technique. 

The best-fit power-law parameters can be found in Table
\ref{power_law}.  We first determined the best-fit parameters of the
LRG auto-correlation function by minimizing the $\chi^2$ function that
accounts for the correlated errors between adjacent bins:
\begin{equation}
\chi^2 = (w-\tilde{w})^T {\rm COV}^{-1}(w-\tilde{w}) \; . 
\label{least-square}
\end{equation} 
$\tilde{w}$ is the model and $w$ is the data vector.  Next, we adopted
the best-fit slope of the LRG auto-correlation function for all
corresponding Mg\,II--LRGs cross-correlation calculations.  For the
LRGs auto-correlation, the errors on the parameters were determined
from all values within $\Delta \chi^2 < 2.3$ (two parameters to fit)
from the minimum. The relative and absolute biases derived from the
ratio of the best-fit power-law amplitude are denoted by the subscript
"RA" in Table 4.  In the cross-correlation cases, the error on the
best-fit amplitude $a$ corresponds to all models with $\Delta
\chi^2<1$ from the minimum $\chi^2$ value.

For the relative bias derived from the direct ratio "DR" of datapoints
in the two-halo regime (equation \ref{dr}), we excluded all negative
datapoints from the calculation. The final error was derived using the
standard error propagation technique and we kept the two most dominant
terms of the expansion. These two terms are at least two orders of
magnitude larger than any other term in the expansion,
\begin{eqnarray}
& &\sigma_{\rm DR}^2 = \sum_i \omega_i^2 \frac{1}{w_{\rm w_{ag,i}}^2}\sigma_{\rm w_{ag},i}^2 +{} \nonumber \\
& & {}+2\sum_i\sum_{j>i}
\omega_i \omega_j {\rm COV}(w_{\rm ag,i},w_{\rm ag,j})\frac{w_{\rm gg,i}w_{\rm gg,j}}{w_{\rm ag,i}w_{\rm ag,j}}
\label{sigma_dr}
\end{eqnarray} 
where the sum runs over the bins $i,j$ with $r_p > 1.1~h^{-1}$
Mpc. $w_{\rm ag,i}$ represents the $i$th bin of the cross-correlation
function and $w_{\rm gg,j}$ is the $j$th bin of the LRGs
auto-correlation function. The relative bias values, calculated using
both methods, are listed in columns (2) and (3) of Table
\ref{bias_table}. We corrected the relative bias values for
photometric redshift errors by using the large scale correction factor
0.90$\pm 0.02$.

\subsection{The absolute bias and  mass scale of absorber hosts}

Since the flux-limited sample of galaxies is not based on a homogeneous 
sample of galaxies and would render the halo occupation analysis 
much more challenging, we limited our absolute bias calculation 
to the volume-limited data. The absolute biases obtained, for the 
direct-ratio case, are $b=1.56\pm0.57$ for the weak Mg\,II sample
and $b=1.09\pm0.47$ for the strong one. Biases derived from the power-law 
technique are found in column (5) of Table \ref{bias_table}. 

We determined the corresponding halo mass in two different ways. In the
first case, we inverted equation (\ref{halo_bias}) to obtain the halo mass for these absorbers. 
These masses are denoted by $\log M_h$, which we refer to as the \emph{bias-inverted} mass.
Using the direct-ratio evaluated according to equation (\ref{dr}), we derived
$\log M_h = 13.0^{+0.4}_{-0.6}$ for the weak absorbers and $\log M_h
= 12.0^{+0.8}_{-6.0}$ for the strong ones. The lower error bar we quote for 
the weak absorbers is arbitrary. All halos with $\log M_h<12$ are consistent with our results. 
This is because the lower error bar on the bias was less than 0.7 
giving us no constraint on the halo mass. Indeed, $b(M)$ has a minimum value of 
$\sim 0.7$ and becomes nearly independent of mass at $\log M_h \lesssim 9$.
The corresponding halo masses using
the power-law method are in column (7) of Table \ref{bias_table}.  
In the second case, we first solved for the minimum halo mass ($M_{min}$) using 
\begin{equation}
\langle b \rangle = n_h^{-1} \int_{M_{min}}^{\infty} d\log M_h \frac{dn(M_h)}{d\log M_h} b(M_h)
\end{equation}
where $\langle b \rangle$ is the absorbers bias, $dn(M_h)/d\log M_h$ is the 
halo mass function and 
\begin{equation}
n_h = \int_{M_{min}}^{\infty} d\log M_h \frac{dn(M_h)}{d\log M_h} \; . 
\end{equation}
We then estimated the mass following 
\begin{equation}
\langle \log M_h \rangle = n_h^{-1}\int_{M_{min}}^{\infty} d\log M_h \frac{dn(M_h)}{d\log M_h} \log M_h \; . 
\label{bw_mass}
\end{equation}
We refer to the mass found using equation (\ref{bw_mass}) as the \emph{bias-weighted} halo mass. 
$M_{min}$ corresponds to the minimum halo mass above which there is a single
absorber per halo. Of course, absorbers are distributed over a wide range of masses and
the covering fraction is likely to vary with halo mass. This method only
gives an approximate answer.  As for the the bias-inverted mass, it is
a characteristic mass  obtained by inverting the
$b(M)$ relation.  Even though these two methods involve different
assumptions, they give similar (within 0.1 dex) results for both
LRG and absorbers masses. We used the bias-inverted masses as our LRG
and absorbers mass estimate.

The relative and absolute biases, calculated with the direct ratio
technique, and their corresponding halo masses are shown in Figure
\ref{bias}. Panel (a) shows the relative bias obtained for the volume-
and flux-limited ($z=0.40-0.70$) samples and allows for a direct
comparison with the \citet{bouche2006a} results; (b) shows the
absolute bias derived from the halo occupation analysis. The
bias-inverted halo masses are shown in panel (c) along with
\citet{bouche2006a} bias-weighted mass estimates.  We did not quote a
lower limit for the halo mass when the lower limit on the absolute
bias is $\lesssim 0.7$. 

We obtained similar relative biases for the flux-limited sample at
$z=0.40-0.70$ as Bouch\'e \etal, who corrected their relative bias by
20\% to account for phototmetric redshifts. This large correction is
expected since these authors included fainter LRGs ($i<21$) in their
analysis that are expected to contain larger photometric redshift
uncertainties.  It is, however, not clear how the authors accounted
for the varying redshift errors with galaxy brightness in their estimate.

In contrast, we applied a 10\% correction factor for the clustering
measurements according to our simulation studies described in \S\
3.2.2.  The halo masses derived for weak/strong Mg\,II are larger than
the ones obtained by \citet{bouche2006a} over similar $W_r(2796)$
intervals.  
The apparent discrepancy may be due to the absolute halo bias of LRGs
included in different analysis.  We calculated $b_{\rm lrg}$ directly
from the LRG clustering signal observed in a volume-limited sample,
whereas \citet{bouche2006a} compared their LRG clustering strength
with previous studies that were carried out using a similar, but not
identical galaxy population.  Additional uncertainties in the
estimated correction factor to account for photometric redshift errors
may also contribute to the differences in our findings.

\section{Discussion}

We have calculated the clustering amplitude of Mg\,II absorbers with
respect to three samples of LRGs.  Using the volume-limited sample, we
have computed the absolute bias and typical halo mass of two
subsamples of Mg\,II absorbers: $W_r(2796)>1.5$ \AA\, and 
$W_r(2796)=1-1.5$ \AA\.  A $\sim 1\sigma$ anti-correlation between
$W_r(2796)$ and mean mass is seen in panel (c) of Figure \ref{bias}.
If a significant anti-correlation signal is confirmed by larger datasets, 
this would imply that weaker Mg\,II absorbers are found to be
more strongly clustered than stronger ones. Our results show that a significant 
fraction of the Mg\,II absorber population of $W_r(2796)=1-1.5$ \AA\
absorbers are found  around massive galaxies
with $ \log M_h < 13.4$, whereas absorbers of $W_r(2796)>1.5$ \AA\ 
are primarily found in $\log M_h < 12.7$ galaxies. Larger datasets for both LRGs and 
Mg\,II would improve the precision on the clustering measurements and the equivalent 
width vs. mass relationship.   
Corresponding galaxy luminosities can
be inferred from the bias-luminosity relation found for SDSS data at
$z\approx 0.1$ (\citealt{tegmark2004a}; see also
\citealt{zehavi2005a}). Assuming $b=1$ for $L_*$-galaxies (which
is reasonable since $L_*$-galaxies are found in halos of mass $\sim
10^{12.3}$ M$_{\odot}$; see \citealt{zheng2007a}), the bias-inferred 
luminosity for weak and strong absorbers are $\approx 4.5 L_*$ and \
$\approx 1.5 L_*$ respectively. Note that these values are obtained 
from the mean bias estimate of equation 15 where more massive galaxies 
(higher luminosity) have higher weights (bias) than less massive (lower 
luminosity) objects. These values should not 
be interpreted as the luminosity of a typical galaxy producing absorbers 
of a given strength. In addition, the relationship between bias and luminosity
is not a one-to-one relation since the bias of galaxies is affected by
the satellites within their dark matter halos. There is also
an expected redshift evolution of the bias-luminosity
relation. However, \citet{zheng2007a} found that there is little
evolution in the halo mass hosting the central galaxies between $z=0$
and $z=1$. These authors found that a typical $L_*$-galaxies reside in
halos only a few times more massive at $z=1$ than at $z=0$.
Therefore, we assume no redshift evolution for the bias-luminosity
relation and used the expression found at $z=0.1$ as a reasonable
guess for our $z=0.5$ sample.

Another important aspect of the results is that the 
Mg\,II-LRG cross-correlation function continues to exhibit
a strong signal (even after correcting for photmetric redshifts) down
to $\sim 0.3\ h^{-1}$ Mpc, indicating that some of the Mg\,II
absorbers and the LRGs share a common dark matter halo.  Here we
discuss the implications of these results.

\subsection{The bias vs $W_r(2796)$ relationship}

The $W_r$ vs.\ mean halo mass relationship found in our analysis
is qualitatively consistent with the previous report by Bouch\'e
\etal\ (2006; see also \citealt{lundgren2009a}). It is a
1-$\sigma$ trend but it argues against the simple notion that more
massive halos might produce stronger absorbers because they contain a
larger volume of gas. Our results show that \emph{weaker} absorbers
are preferentially found in more massive halos. 

Bouch\'e \etal\ (2006) measured the Mg\,II-LRG cross-correlation
function using 1806 Mg\,II absorbers of $W_r(2796)>0.3$ \AA\ and
250,000 LRGs of $i'<21$ at $z=0.35-0.8$.  They found that 
Mg\,II absorbers of $W_r(2796)\apll 1$ \AA\ appeared to be more
strongly clustered than the $W_r(2796)\apg 2$ \AA\ ones by
nearly a factor of two.  The authors attributed the observed
anti-correlation to a starburst outflow origin for absorbers
with $W_r(2796)\apg 2$ \AA, in order to explain the on-average lower
halo mass of these absorbers.  

However, our analysis shows that absorbers with $W_r(2796)\apg 1.5$ 
are essentially unbiased with respect to dark matter ($b=1$). 
This indicates that the
halo population probed by these absorbers is consistent with a
random, unbiased sample of dark matter halos, and does not favor a
specific sub-population such as starbursting systems.  For
absorbers of $W_r(2796)<1.5$ \AA, the mean halo bias was found to be
still higher.  This large mass scale is at odds with previous
findings that Mg\,II absorbers of $W_r(2796)=0.3-1$ \AA\ are
associated with $L_*$-type galaxies (e.g.,\ Steidel \etal\ 1994).

To understand the physical mechanisms that could explain a 
$W_r(2796)$ vs.\ clustering amplitude anti-correlation,
\citealt{tinker2008a} (hereafter TC08) developed a halo occupation
model that constrains the cold gas content of dark matter halos based
on the observed number density and clustering amplitudes of the Mg\,II
absorbers.  In the TC08 model, Mg\,II absorbers serve as a
representative tracer of cool gas ($T\sim 10^4$ K) in dark matter
halos, and the observed anti-correlation arises as a result of an
elevated clustering amplitude of $W_r(2796)=0.3-1.5$ \AA\ absorbers
due to the contributions of residual cold gas in high-mass halos
($M_h>10^{13}\,\hmsol$).  These massive halos are rare and are likely
missed in small samples (e.g., Steidel \etal\ 1994).  The halo
occupation model represents the first empirical constraint of the cold
gas content across the full spectrum of dark matter halos and provides
additional information for models of the growth of gaseous halos.

\subsection{Presence of cool gas in massive halos}

From Figure \ref{everything}, it is worth noting the strong clustering
of Mg\,II absorbers for all three LRG samples.  Indeed, the
clustering strength is comparable to the LRGs auto-correlation signal
for the most inner bin ($r_p = 0.31~h^{-1}$ Mpc). In physical units, 
this bin corresponds to 0.21$~h^{-1}$ Mpc. For the measured
clustering scale, the typical mass scale of LRGs is roughly
$\rm{log}(\rm{M}_h/\rm{M}_{\odot}) \approx 13.2$, implying a
typical virial radius of $R_{\rm vir} \sim 0.35~h^{-1}$
Mpc. The virial radius is larger than the physical separation
probed by the most inner bin. The strong cross-correlation 
signal is thus indicative of the presence of cool gas well 
inside the virial radius of massive galaxies.

To further investigate the presence of cool gas in massive halos, we
examined the effects that a varying cold gas covering fraction
($\kappa$) in LRG halos would have on the cross-correlation signal. 
To do this, we constructed a mock LRG catalog based on our halo occupation 
distribution fits to the volume-limited sample (see bottom left panel of Figure 4). Using 
the best-fit halo occupation function, we populated the halos 
identified in a z=0.5 output of an N-body simulation. This simulation 
is smaller in volume (400 $h^{-1}$ Mpc on a side)
than the simulation used to test the photometric redshift errors, in order to probe lower-mass
halos. Details about this simulation can be found in \citet{tinker2007a}.
The cosmology of this simulation differs from our fiducial cosmology (WMAP1 
vs WMAP5), so the large-scale bias of the LRG auto-correlation function 
is lower in comparison to the data, but the one-halo clustering 
is a good match to the data. 

We then simulated a mock Mg\,II absorber catalog by selecting random
sightlines.  Every halo of $M_h\ge 10^{12} \hmsol$ was allowed to
produce a mock Mg\,II absorber if the impact parameter was less
than the virial radius, whether or not it contains a mock LRG. We then 
measured the cross-correlation between mock absorbers and mock LRGs 
for different values of $\kappa$ following different recipes.  First,
all halos of containing an LRG yielded an absorber if intersected by
a sightline.  Namely, all LRG halos have a gas covering fraction of $\kappa=1$.
Then, we varied the covering fraction of Mg\,II in the mock LRG halos.

\begin{figure*}
\centerline{
\includegraphics[angle=-90,scale=0.80]{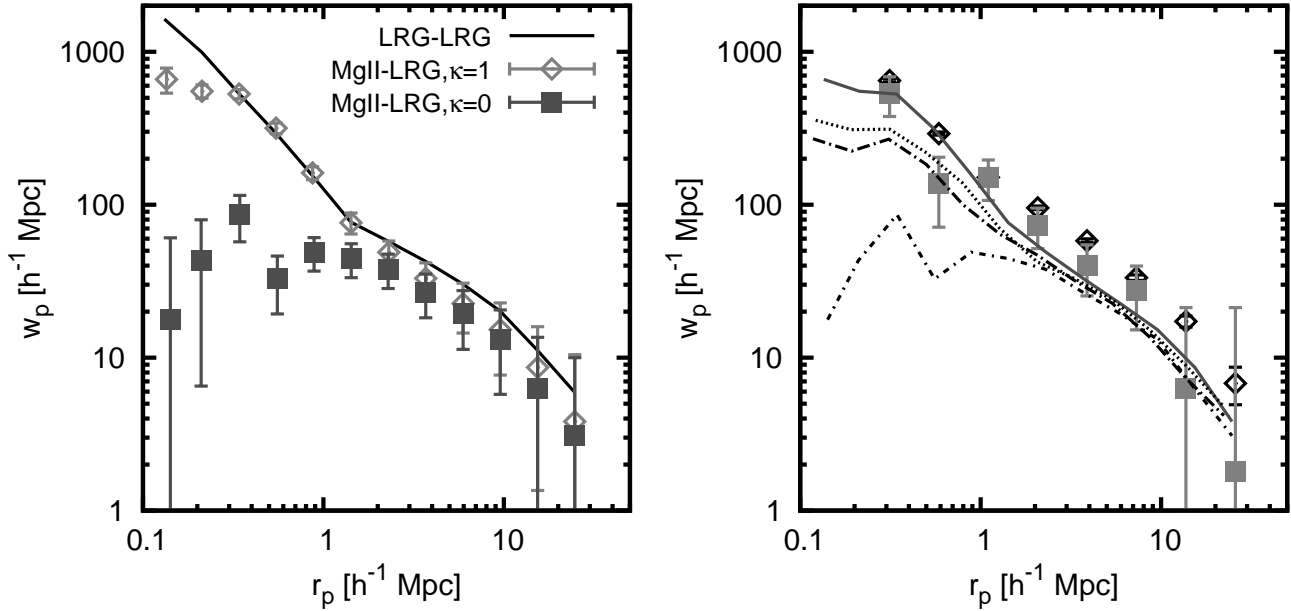}
}
\caption{\emph{Left panel}: Cross- and auto-correlation results for the 400 $h^{-1}$ Mpc box. 
The \emph{solid line} shows the auto-correlation of the mock LRG sample. The \emph{diamond} points 
  show the cross-correlation results for which $\kappa=1$ for all halos above $10^{12}~h^{-1}$ 
M$_{\odot}$. The \emph{square} points represent the case of no cold gas in the halos of the mock LRGs and 
$\kappa=1$ for all remaining halos $>10^{12}~h^{-1}$ M$_{\odot}$. \emph{Right panel}: the cross- and 
auto-correlation results for the volume-limited sample of LRGs ($W_r(2796)=1-5$ \AA) are shown in 
gray and black points respectively. The four curves correspond to different values of $\kappa$ for the mock LRGs.
From bottom to top, $\kappa=0,0.33,0.5,1$.  
}
\label{cool_gas}
\end{figure*}

The results for the two limiting cases of $\kappa=0$ and $\kappa=1$
are shown in the left panel of Figure \ref{cool_gas} along with the
mock LRG auto-correlation function calculated from the box. The right
panel shows four curves corresponding to four different $\kappa$
values on top of our volume-limited cross-correlation and
auto-correlation measurements for $W_r(2796)=1-5$ \AA\ absorbers. 
The cross-correlation function is a probe of the relative covering 
fraction of LRG halos with respect to lower mass halos. Lowering $\kappa$ by the 
same amount for all halos, LRGs and $L_*$ alike, does not change the 
resulting cross-correlation signal. The results from Figure 7 imply 
that the covering fraction of LRG-hosting halos must be comparable 
to that of halos that contain $L_*$-galaxies at their centers. 

The presence of cold gas in massive halos has been a debated subject
in recent numerical simulation studies. 
In a series of high-resolution SPH simulations,
\citet{keres2008a} and \citet{brooks2008a} examined the temperature
history of gas accreted onto dark matter halos. They found that most
of the baryonic mass is acquired through filamentary cold mode
accretion that is never shock heated to the virial temperature for halos of 
$\le 10^{12}$ \hmsol. \citet{keres2008a} found that these cold flows 
are not present in massive halos typically hosting LRGs. 
At high redshift, cold flows may penetrate inside the virial shock 
of $10^{13}~h^{-1}$ M$_{\odot}$, but this effect is highly redshift dependent, 
and is not likely to yield to cross-correlation functions seen in Figure 4. 

Other mechanisms such as thermal instability could generate pockets of
cold gas inside a hot medium (e.g.,\ \citealt{mo1996a}). \citet{maller2004a} 
showed that the hot gas is thermally unstable and
prone to fragmentation. They also show that cooling proceeds via the
formation of cold $10^{4}$ K clouds in pressure equilibrium with the
hot halo gas. For a Milky-Way-size system, cool clouds of mass $\sim 5
\times 10^6$ $\rm{M}_{\odot}$ are expected to extend up to $\sim 150$
kpc from the galactic center and survive for several Gyrs. \citet{kaufmann2008a} showed
that cloud formation is viable in $M_*$ halos, but needs to be extended to higher mass. In a
similar argument developed by \citet{mo1996a}, a two-phase medium in
pressure equilibrium was used to explain observations of Lyman limit
systems. This model also makes predictions about the presence of C\,IV
around low-mass galaxies and at large impact parameters of massive
galaxies.  These predictions were partially confirmed later by
\citet{chen2001b} who found C\,IV in galaxies of different
morphologies and luminosities. These authors also observed the sharp
boundary in the $W_r$-(projected separation) plane that was also
predicted. \citet{mo1996a} attributed the presence of Mg\,II absorbers
to cold pockets of photoionized gas in halos around massive galaxies.
The observed strong Mg\,II-LRGs cross-correlation signal on scales
smaller than the virial radii of halos hosting LRGs indicates the
presence of cold gas is more common around massive galaxies than
previously thought.

On the observational side, the detections of cool gas in 
group size halos is uncertain. The challenges lie 
in the limited sensitivities available to detect HI gas via 21-cm 
observations (e.g.,\ \citealt{verdes-montenegro2001a,verdes-montenegro2006a}). 
\citet{verdes-montenegro2006a} reported detections of HI column density 
down to $\sim 10^{19}$ cm$^{-2}$ in some of the Hickson Compact Groups 
\citep{hickson1982a}. Observations are not sensitive enough to 
probe the low HI column density environment yet. 

The difficulty in finding cool gas around groups using 21-cm observations 
underscores the powerful application of QSO absorption-line studies.
We are currently conducting a follow-up imaging and spectroscopy 
campaign to study the cool gas content of individual LRG halos 
(Gauthier et al. 2009 in preparation).

\subsection{Future prospects : DR7 Mg\,II database \& HOD modeling}

Figure 7 demonstrates how a halo occupation approach to the galaxy-absorber
cross-correlation function at both large and small scales can put constraints 
on the covering fraction of cold gas in LRG-hosting dark matter halos. 
In a forthcoming paper, we will address the detailed halo occupation
distribution modeling of the Mg\,II absorber environment with a
particular focus on the one-halo term. This will be achieved by using
the SDSS DR7 Mg\,II absorber catalog.  This catalog, currently in
preparation (Prochaska et al.\ 2009), will more than
double the number of absorbers from the current DR5 sample. It will
allow us to probe smaller projected separations 
and improve the clustering measurements at $r_p < 1~h^{-1}$ Mpc.  

\acknowledgments

We thank C.\ Blake, I.\ Zehavi, and A. Kravtsov for useful discussions and 
the anonymous referee for useful comments 
that improved the draft. We are grateful to the NYU-VAGC team 
for generating the SDSS survey masks and
making them available through their website and to the MANGLE team for
making their software publicly available.  We thank J.X. Prochaska for
providing an updated SDSS DR5 Mg\,II catalog, and H.\ Oyaizu for
providing the electronic table necessary for producing Figure 3 in
this paper.  HWC acknowledges partial support from NASA Long Term
Space Astrophysics grant NNG06GC36G and an NSF grant AST-0607510.

\bibliographystyle{apj}
\bibliography{ms11ref}

\clearpage

\end{document}